\documentclass[useAMS,usenatbib]{mn2e}
\usepackage{graphicx}
\usepackage{epstopdf}

\title[Where are the missing gamma ray burst redshifts?]{Where are the missing gamma ray burst redshifts?}
\author[D. M. Coward]{D. M. Coward,$^{1}$\thanks{E-mail:
coward@physics.uwa.edu.au} D. Guetta$^{2}$, R. R. Burman$^{1}$ and A. Imerito$^{1}$\\
$^{1}$School of Physics, University of Western Australia, M013, Crawley WA 6009, Australia\\
$^{2}$INAF-Osservatorio Astronomico di Roma, Monteporzio
Catone (Roma), Italy.}
\begin{document}

\date{Accepted Received ; in original form }

\pagerange{\pageref{firstpage}--\pageref{lastpage}} \pubyear{2007}

\maketitle

\label{firstpage}

\begin{abstract}
In the redshift range $z = 0-1$, the gamma ray burst (GRB) redshift distribution should increase rapidly because of  increasing differential volume sizes and strong evolution in the star formation rate. This feature is not observed in the {\it Swift} redshift distribution and to account for this discrepancy, a dominant bias, independent of the {\it Swift} sensitivity, is required. Furthermore, despite rapid localization, about 50\% of {\it Swift} and pre-{\it Swift} GRBs do not have an observed optical afterglow and  60-70\% of GRBs are lacking redshifts.  We employ a heuristic technique to extract this redshift bias using 69 GRBs localized by {\it Swift} with redshifts determined from absorption or emission spectroscopy. For the {\it Swift} and {\it HETE}+{\it BeppoSAX} redshift distributions, the best model fit to the bias at $z<1$ implies that if GRB rate evolution follows the SFR, the bias cancels this rate increase. We find that the same bias is affecting both {\it Swift} and {\it HETE}+{\it BeppoSAX} measurements similarly at $z<1$. Using a bias model constrained at a 98\% KS probability, we find that 72\% of GRBs at $z<2$ will not have measurable redshifts and about 55\% at $z>2$. To achieve this high KS probability requires increasing the GRB rate density at small $z$ compared to the high-$z$ rate. This provides further evidence for a low-luminosity population of GRBs that are observed in only a small volume because of their faintness.
\end{abstract}


\section{Introduction}
Because GRBs are the brightest transient cosmological events measurable, they are a unique probe to the high-$z$ Universe and provide an opportunity to study their local interstellar environment, host galaxies and the intergalactic medium. Multi-wavelength observations of GRBs have confirmed that a significant fraction of long GRBs \footnote {Hereafter GRB refers to bursts classified as long.}  are associated with the collapse of short-lived massive stars \citep{hjorth03,stan03}. This provides strong motivation to use the measured GRB redshift distribution as a tool to constrain the star formation rate (SFR) and GRB rate evolution. However, it is clear that the relation between the evolving SFR and the GRB redshift distribution is more complex than originally assumed. 

Studies of the host galaxies of supernovae reveal that they are more probable in brighter, massive spirals, while GRBs are more typically in smaller irregular host types \citep{fruch06}. A possible explanation for the difference in host galaxy types is a preference for GRBs to form from relatively lower-metallicity core-collapse supernovae \citep{yoon05}.  This argument is compatible with the `collapsar' model \citep{woos93}, which postulates that massive GRB progenitors may be preferred in lower-metallicity environments. The inferred metal abundances in the GRB host galaxies are on average smaller than those found in ultra-luminous dusty galaxies (e.g. Le Floc'h et al. 2006).  Although the SFR is lower in the GRB host galaxies, it still may comprise about 2/3 of the global SFR. Hence, the observed GRB redshift distribution is possibly tracking the SFR in the faint end of the galaxy luminosity distribution. 

Up to 2007 November, more than 80 GRB redshifts have been measured from rapid localization by the {\it Swift} satellite and follow-up spectroscopy of the optical/NIR afterglow and host galaxies by large ground-based telescopes.  Optical/NIR afterglows have been found for nearly $60\%$ of {\it Swift} localized GRBs but only about $30\%$ of {\it Swift} GRBs have measured redshifts, implying a very incomplete sample.  The reasons why a significant fraction of localized GRBs have no detectable afterglow is an ongoing controversy. It is known that GRB afterglow radiation is subject to several absorption processes in the optical band \citep{gou04}. The presence of dust and gas, though, does not obscure gamma rays, irrespective of whether this material is within the host galaxies or in the intergalactic medium \citep{Blain00}.  

\section{Optically dark GRBs}
GRB optical afterglows may be dark due to obscuring dust, very high redshifts or rapidly decaying transients, a phenomena which is commonly referred to as a dark burst--see \cite{MR03} and references therein. A subclass of `intrinsically dark' GRB afterglows has been argued against by \cite{Mz05}. They show that selection effects are strong enough to explain the proportion of GRBs deemed to be optically dark. Such `dark' bursts may be a result of dust extinction in the local ISM of the host along the line of sight. Evidence for this is shown in a recent study of the host galaxy of GRB 030115 \citep{lev06}. Its optical afterglow was fainter than many upper limits for other bursts, suggesting that without early NIR observations it would have been classified as a dark burst. Both the colour and optical magnitude of the afterglow are probably affected by dust extinction and indicate that at least some optical afterglows are very faint owing to dust along the line of sight. Furthermore, the GRB host galaxies are on average smaller and fainter than supernova host galaxies, so it is more difficult to identify many GRB hosts. 

In another study, \cite{rom2006}, used very early observations of {\it Swift} GRBs to investigate the most probable cause of optically dark GRBs. They find that $\sim 25\%$ of the bursts in their sample are extincted by Galactic dust, $\sim 25\%$ are obscured by absorption in the immediate GRB environment and $\sim 30\%$ are most likely lost by Ly-$\alpha$ blanketing and absorption at high redshift. So it is highly likely that the dark bursts result from a combination of extinction factors relating to the GRB environment, host galaxy type and redshift. \cite{schad2007} estimate that GRBs without an afterglow detected in $\lambda> 5500$ Angstrom have average optical extinctions eight times those observed in the optically bright population of GRBs. They point out that extinction could account for the non-detection of GRB afterglows by the UVOT onboard Swift.

Given the incompleteness of the {\it Swift} GRB redshift sample, one strategy for using the current redshift distribution as a high-$z$ probe is to remove bursts from the sample where optical afterglow observation was sub-optimal. \cite{jakob06} and \cite{T2007} used a set of criteria to select a sub-sample of GRBs based on Galactic foreground extinction and Sun and Moon positions relative to the burst. Despite this culling they find that the redshift completeness of {\it Swift} GRBs could be increased only to about 50-60\%, implying that a combination of other selection effects are severely affecting the sample.

In this study, we use a different approach, that focusses on investigating how the redshift sample is modified by a combination of unkown selection biases. We show how a global selection function can be extracted from GRB redshift data selected with minimal assumptions. We do not attempt to model the selection function based on strong physical arguments, because of the large uncertainty in the multiple biases that comprise the selection function. Regardless of this limitation, the form of the selection function extracted from our analysis provides a means to estimate how the dominant biases may be modifying the redshift distribution. To achieve this aim we employ a subset comprised of 69 {\it Swift} redshifts obtained from well-localized GRBs with $T_{90}>2$s and accurate spectroscopic redshifts. We have not included photometric redshifts or upper and lower limits for redshifts. The sample includes the controversial GRBs 060614 ($z=0.12$) and 060505 ($z=0.089$), which were long duration events without accompanying supernovae.  

The goal of this study is twofold: firstly to constrain a model for the combined biases using a heuristic approach with minimal assumptions. Nonetheless, we do include a very crude model for the so-called `redshift desert' at $z\approx1-2$ -- a region where it is difficult to measure redshifts because of the lack of observable strong emission or absorption lines \citep{fiore07}. We point out that this bias is restricted to those GRBs with optical afterglows--i.e. 50-60\% of the total {\it Swift} GRB sample, implying a combination of other astrophysical and observational selection effects modify the sample. To account for the remaining  missing fraction of redshifts, we employ a parametrized function that scales the fraction of GRB afterglow detections at $z=0.1-6$.

Secondly, we will examine if the fractional contributions of individual selection biases (e.g. Roming et al. 2006) are reconcilable with the global selection function extracted in this study. The elegance of this technique is that the selection function tracks the evolution of the dominant biases in different redshift ranges. The selection function extracted in our analysis is constrained using a KS test applied to the chosen subset of GRB reshifts and a bias-corrected probability distribution model. 
This is a first step toward using the distribution of  `missing' GRB redshifts as a potential new probe of the evolution of GRB environments. 

\section{The GRB redshift selection function}

An inspection of the {\it Swift} redshift distribution provides an insight into the form of such a function. Firstly, in the redshift range $0-1$, the distribution should increase rapidly because of increasing differential volume sizes. Assuming that GRB rate evolution follow the global SFR in a 1:1 manner, recent observation-based models of the SFR at $z<1$  imply a rapid increase in the GRB redshift distribution in the same range. This feature is absent from the present distribution. Secondly, the distribution shows a distinctive drop in numbers at $z \approx 1-2$, and an increase at $z>2$. Figure 1 plots the {\it Swift} normalized redshift distribution (shaded region), along with a model distribution. 

 To model a complete GRB redshift selection function requires knowledge of the component biases. One is the sensitivity of the satellite detector assuming a distribution of  source luminosities. For {\it Swift}, there are also complex detector triggering algorithms that may introduce a bias, but these effects have not been modelled completely and we do not attempt to account for them here. Since redshift measurement depends on obtaining optical emission or absorption spectra, it is clear that any selection of optical afterglows will affect the redshift distribution. Because about 50\% of GRBs are optically dark, there is a dominant selection effect operating on the optical afterglow distribution, as discussed above. 

Of the GRBs that do have observed optical afterglows, only about 60\% have measured spectroscopic redshifts. It is apparent that other biases related to to the acquisition of spectroscopic redshifts may play a part. The  `redshift desert', $z\approx1-2$, is a region where it is difficult to measure redshifts because of the lack of strong emission or absorption lines covered by optical spectrometers \citep{fiore07}. Strong emission lines such as H$\alpha$, H$\beta$, OIII and OII go out of the observed range at $z\approx1.1$ while Lyman-$\alpha$ enters the range at $z\approx2.1$. The strong absorption feature of the MgII ($\lambda=2796,2803$) doublet is contaminated by telluric features at $z\ga1.5$.

The magnitudes of the individual biases from these effects are expected to be very different. For instance, dust extinction may well be a dominant effect given that a significant fraction of GRBs are optically dark. Our approach is to estimate a `global' selection function by weighting the redshift sample by the flux-limited sensitivity of the detector and volume-number dependence from the redshift sample. The geometric component is well known but the flux-limited sensitivity is less so because of the uncertainty in the GRB luminosity function. 

For self consistency we assume a single luminosity function (LF) that encompasses the high luminosity (HL) GRBs and the so-called anomalous low-luminosity (LL) GRBs, those with luminosities  $<10^{49}$ ergs s$^{-1}$, detected by {\it HETE}+{\it BeppoSAX} and {\it Swift}. We adopt the double power-law LF, $\phi (L)$, from \cite{guetta05}, with slopes $-0.1$ and $-2$, with luminosity limits that encompass the extremely LL GRB 980405: $(L_{\rm max},  L_{\rm min}) = (3 \times 10^{53}, 3 \times 10^{46})$ ergs  s$^{-1}$. The fraction of detectable GRBs, or flux-limited selection function (those observed with peak flux  $>F_{\rm lim}$), is
\begin{equation}\label{grbrate}
\psi_{\rm flux}(z)= \int_{L_{\rm lim}(F_{\rm lim},z)}^{\infty}\phi(L_{\rm iso}) {\rm d}L_{\rm iso} \;,
\end{equation}
where $L_{\rm iso}$ is an isotropic equivalent luminosity and $L_{\rm lim}(F_{\rm lim},z)$ is obtained by solving $F_{\rm lim}(L_{\rm lim},z) = L_{\rm lim} / (4 \pi d^2_{\mathrm L}(z))$, with $d_{\mathrm L}(z)$ denoting the luminosity distance of the burst. The flux sensitivity limit, $F_{\rm lim}$, of the detector BAT onboard {\it Swift} is about $1\times 10^{-8}$ ergs cm$^{-2}$ s$^{-1}$ and for a combined {\it HETE}+{\it BeppoSAX} sensitivity we take a mean of $4 \times 10^{-8}$ ergs cm$^{-2}$ s$^{-1}$, approximated from \cite{Band06}.

\subsection{The GRB redshift probability distribution function}\label{prob}
The main redshift distribution biases are represented as the product $\psi_{\rm flux}(z)\psi_{\rm redshift}(z)$ of two dimensionless selection functions, where $\psi_{\rm redshift}(z)$ describes an evolving bias comprised of both spectroscopic and astrophysical components; $\psi_{\rm redshift}(z)$ is uncertain, but plays a fundamental role in shaping the redshift probability distribution function --- see  \cite{Cow07} for a more detailed derivation. The GRB probability distribution function is expressed as: 
\begin{equation}\label{pdf1}
P(z) = {\rm N}^{-1}\psi_{\rm flux}(z)\psi_{\rm redshift}(z)e(z)\frac{{\mathrm d}V(z)}{{\mathrm d}z}\frac{1}{(1+z)}\;,
\end {equation}
with normalizing factor
\begin{equation}\label{norm}
{\rm N} \equiv \int_{0}^{z_{\rm
max}}\psi_{\rm flux}(z)\psi_{\rm redshift}(z)e(z)\frac{{\mathrm d}V(z)}{{\mathrm d}z}\frac{{\rm d}z}{(1+z)} \;.
\end{equation}
The $(1+z)^{-1}$ accounts for cosmological time dilation, d$V(z)$ is the volume element in a flat-$\Lambda$ cosmology ($\Omega_{\mathrm M}=0.3$, $\Omega_{\mathrm \Lambda}=0.7$) with $H_0=70$ km s$^{-1}$ Mpc$^{-1}$ and ${z_{\rm max}}$ is the maximum observable redshift. We assume that the dimensionless GRB rate evolution factor, $e(z)$, follows the evolving SFR density, $\dot\rho_{*}(z)$. 

\citet{hb06} have constrained the SFR by fitting to ultraviolet and far-infrared data obtained up to 2006. When these data are taken together, they show that it is possible to constrain the SFR out to $z \approx 6$, with especially tight constraints for $z < 1$. Equation (\ref{sfr}) shows the functional form they employed in their fit to the SFR:
\begin{equation}
e(z)  \equiv \frac{\dot\rho_{*}(z)}{\dot\rho_{*}(0)}= \frac{1 + a_1 z}{1+ (z/a_2)^{a_3}}
\, \label{sfr}
\end{equation}
where the evolution function $e(z)$ is defined so that $e(0)=1$. We use the best fit parameters for $e(z)$ based on this SFR model: $(a_1,a_2,a_3)=(7.6,0.0059,4.3)$.

Combining $ (1+z)^{-1}dV/dz$ with the flux-limited satellite selection bias, $\psi_{\rm flux}(z)$, yields a resulting distribution that is strongly peaked at $z=1-2$. Figure 1 plots the distribution described by equation (\ref{pdf1}) assuming $\psi_{\rm redshift}(z)$ is constant (i.e. no bias correction). The {\it Swift} GRB redshift distribution, comprised of 69 spectroscopic redshifts, is clearly different in shape, highlighting a redshift dependent selection bias other than the flux sensitivity of the satellite. The most obvious features are the lack of numbers at $z \approx 1-2$ and the relatively high numbers at $z>3$ compared to $z= 0-1$.
\begin{figure} 
\includegraphics[scale=0.66]{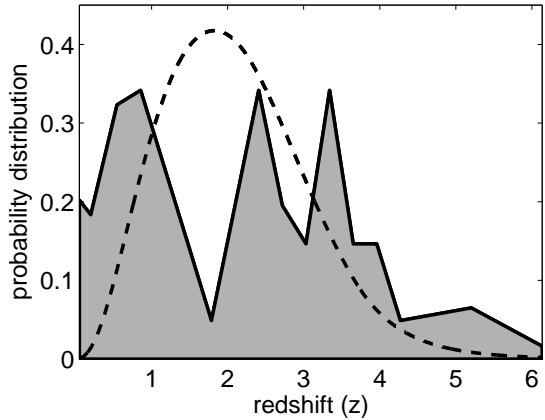}
\caption{Plot of the normalized distribution of 69 spectroscopically measured redshifts obtained following {\it Swift} localizations, shaded, and probability distribution model for the same distribution, with no correction for biases. The model assumes a GRB rate evolution that follows the SFR model of \citet{hb06}. It peaks at $z=1-2$, where the distribution tracks the geometric increase in numbers. At $z>2$, the rapid fall-off is caused by the reduction in flux sensitivity of {\it Swift} and decreasing cosmological volume shells.} \label{fig_pdf} 
\end{figure}

\section{ANALYSIS TECHNIQUE}
Because the probability distribution function (\ref{pdf1}) is separated, the combination of flux-limited selection bias and geometric components,  $\psi_{\rm flux}dV/dz$, can be used to weight the observed redshift distribution. Furthermore, because the SFR is reasonably constrained at $z=0-1$, and assuming that $e(z)$ follows the SFR, this can also be used to weight the observed redshift distribution. If $\psi_{\rm redshift}(z)$ evolves with redshift, the residuals that remain after weighting the observed distribution will not be uniformly distributed. The aim of the analysis is to construct models for $\psi_{\rm redshift}(z)$ that minimize the variance of the residuals after weighting the observed redshift distribution. As a further constraint, we calculate cumulative distributions for the observed and bias-corrected model probability distributions and apply a KS test to quantify their compatibility. 

A further observational constraint on $\psi_{\rm redshift}(z)$ is the fraction of measured redshifts out of the total GRBs localized by {\it Swift}, namely about $30\%$. We approximate this function for the redshift detection efficiency so that it scales according to the fraction of GRBs with measured redshifts. The redshifts are split into 3 ranges for the {\it Swift} and {\it HETE}+{\it BeppoSAX} samples to help identify the effect of different redshift biases. The first, $z=0-0.1$, is dominated by LL GRBs that occur at a possibly higher rate than the HL GRBs.  In the following analysis we constrain the relative rates of the LL bursts to the HL high-$z$ population in this small redshift range. 



The rapid decrease in redshift numbers at $z < 1$ from
the observed distribution is similar to the inverse of the slope
of $e(z)$ in the same range. It is difficult to determine if there
is a direct relationship between this bias and the SFR.
Dust obscuration may play an important role in modifying
the afterglow detection rate, but given the substantial uncertainty
in the links between dust obscuration, SFR evolution
and GRB redshift detection, we take a heuristic approach.

\begin{figure} 
\includegraphics[scale=0.66]{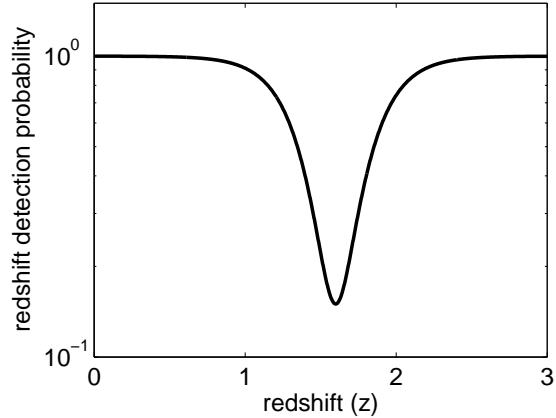}
\caption{Plot of the the toy model for the `redshift desert', equation (\ref{des}), assuming parameters $(A,B)=(0.15,3.00)$ and centred on $z=1.6$. The effect of this suppression on redshift determination is negligible outside of $z = 1-2$.} \label{desert}
\end{figure}

We include a crude model for the redshift desert. From figure 3 of Bloom (2003), the probability of obtaining prominent emission and absorption lines at $1.5 < z < 2$ is reduced by at least 50\%. The same figure shows that at $2 < z < 3.5$ the probability of detection increases significantly. We employ a parametrized toy model for the redshift desert that scales the probability of detection at $1<z<2$. It is essentially unity outside of this redshift range and decreases to a minimum at $z=1.6$:

\begin{equation}\label{des}
\psi_{\rm d}(z;A,B)= 1-(1-A)\mathrm{sech}^{2}[B(z-1.6]\;,
\end{equation}
where $A$ is the minimum value and $B$ defines the width of the desert. Figure \ref{desert} plots $\psi_{\rm d}(z)$ with $(A, B)=(0.15,3.00)$. In this model, redshift determination is affected only in $1< z < 2$, where the desert is prominent. 
  
We separate the contributions of  $\psi_{\rm redshift}(z)$ into two parts. The first, comprised of LL GRBs, allows for a relatively higher rate density than the HL GRBs.  At $z <0.1$,  the $R_{LL}$ parameter scales the GRB rate density of LL GRBs, all observed at small redshift, to the HL high-$z$ bursts:  
\begin{equation} \label{red1a}
\psi_{\rm LL}(z) = 
R_{LL} \; \textrm{for $z \la 0.1 $}\;.
\end{equation} 
For $z>0.1$,  $\psi_{\rm redshift}(z)$ is constrained by the {\it Swift} HL GRB redshift distribution, and assumes the form $\psi_{\rm d}(z)/e(z)$. Two broad redshift ranges are defined, the first one is dominated by $\psi_{\rm d}(z)$ and $1/e(z)$, the second by $1/e(z)$ and a redshift dependent power law of the form $z^{\gamma}$:
\begin{equation} \label{red1b}
\psi_{\rm HL}(z) = \left\{ \begin{array}{ll}
\alpha \psi_{\rm d}(z)/e(z) & \textrm{for $0 < z \la1.85$}\\
\beta z^{\gamma} \psi_{\rm d}(z)/e(z) & \textrm{ $z \ga 1.85$}\;
\end{array} \right.
\end{equation} 
where $z=1.85$ is chosen to ensure that the model joins smoothly at the boundary between the two redshift ranges. The parameters $\alpha$, $\beta$ and $\gamma$ are constrained by finding the highest KS probability using the weighting technique described above. The total bias function is defined as:
\begin{equation}\label{red}
 \psi_{\rm redshift}(z) =  \psi_{\rm LL}(z)+ \psi_{\rm HL}(z)\;.
\end{equation}


We apply the same technique and parametrized form
to the sample of 33 GRB redshifts localized by
{\it HETE}+{\it BeppoSAX}, assuming the mean flux threshold of
these detectors. \cite{fiore07} point out that the median delay time of optical follow-up for  {\it Swift} and {\it HETE}+{\it BeppoSAX} GRBs is 15 minutes and $3.5 -14$ hours respectively. This results in a smaller proportion of faint afterglows triggered from {\it HETE}+{\it BeppoSAX}, so that a bias is introduced against measuring redshifts from high-$z$ GRBs -- i.e. those with faint optical afterglows. To account for this bias, we employ a power law of the form $P_T = z^{-\delta}$  at $z > 1$, and constrain ${\delta}$ using the KS test.

Given that  $\psi_{\rm redshift}$ can be constrained, the fraction of
HL GRB redshifts that are not observed as a function of redshift,
$F_{dark}(z)$, can be calculated from
\begin{equation}\label{fdark}
F_{dark}(z) =1- z^{-1}\int_0^{z}\psi_{\rm HL}(z') dz'\;\; \textrm{for $z \le 5$}\;.
\end{equation} 

We assume that $F_{dark}(z)$ represents the contributions from optically dark GRBs and those with observed afterglows, but with no measured redshift. From the {\it Swift} sample we have selected, the total fraction comprised of both contributions equates to $60-70\%$. It is interesting to consider  the affect on $F_{dark}(z)$ of using a more selective sample. For example, a sample excluding any bursts located near the Sun or Moon increases the completeness of the GRB redshift sample. The effect of fractional completeness on $F_{dark}(z)$ is to scale linearly  the function but not change its shape unless redshifts are omitted from the sample. Because the redshift sample we employ is statistically small, we have chosen not to reduce the sample size further. In fact, doing so would be contrary to our main goal, which is to study how the biases modify the redshift distribution.

\section{Results}
\begin{figure} 
\includegraphics[scale=0.65]{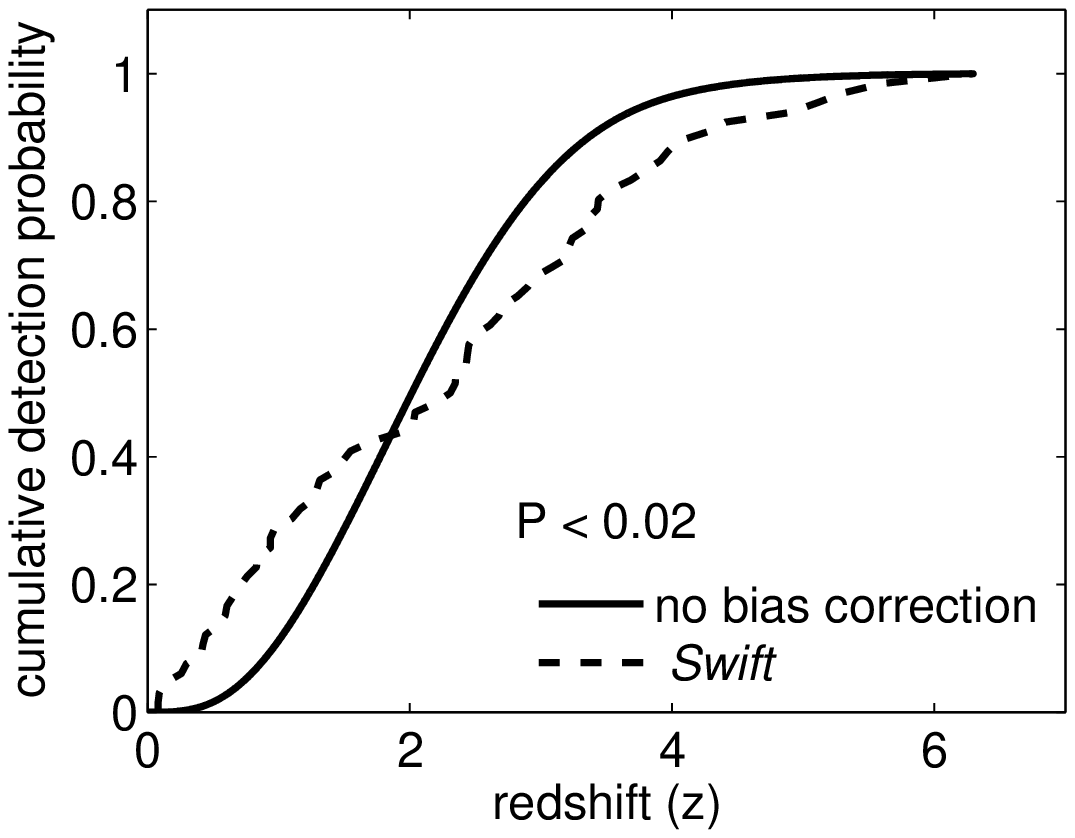}
\includegraphics[scale=0.65]{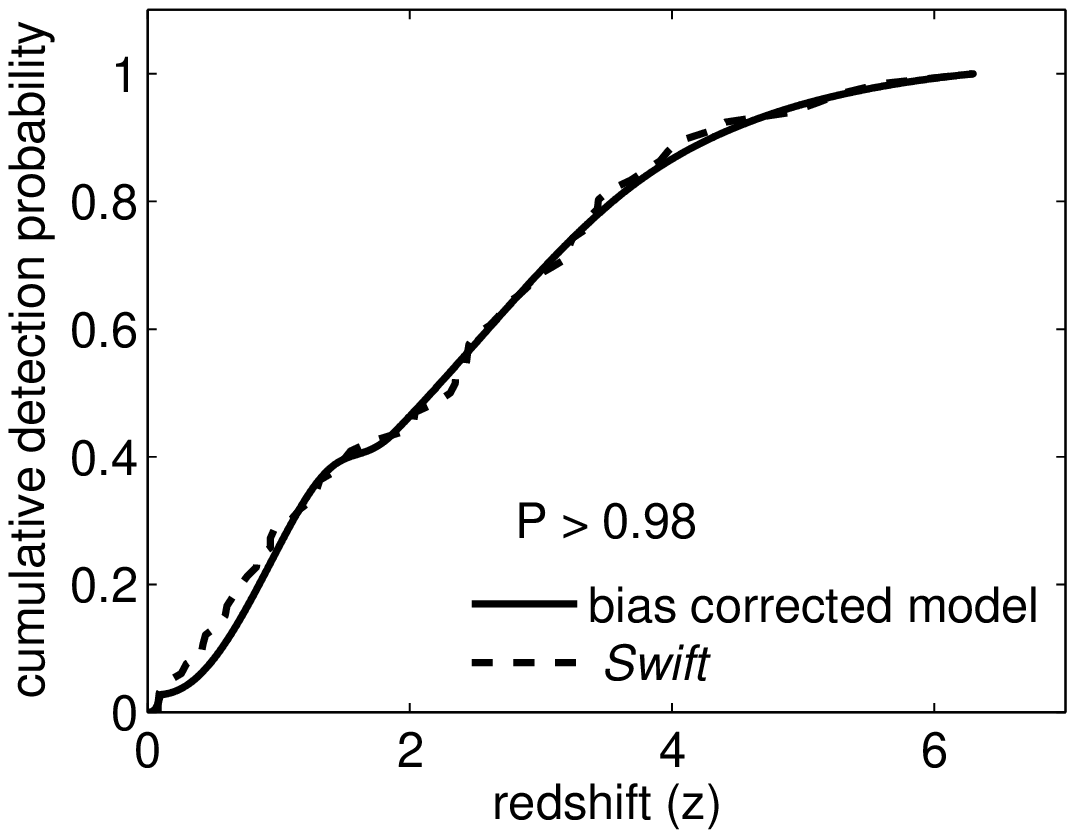}
\caption{Top panel -- plot of the cumulative distribution of {\it Swift} GRB
redshifts and a model distribution not corrected for biases. Lower panel -- the best fit model that includes bias correction using parameter values for $\psi_{\rm redshift}(z)$ of ($R_{LL}$, $\alpha$, $\beta$, $\gamma$) = (40, 1.7, 0.38, 1.5). The corresponding KS probabilities are stated.} \label{figsw} 
\end{figure}

\begin{figure} 
\includegraphics[scale=0.65]{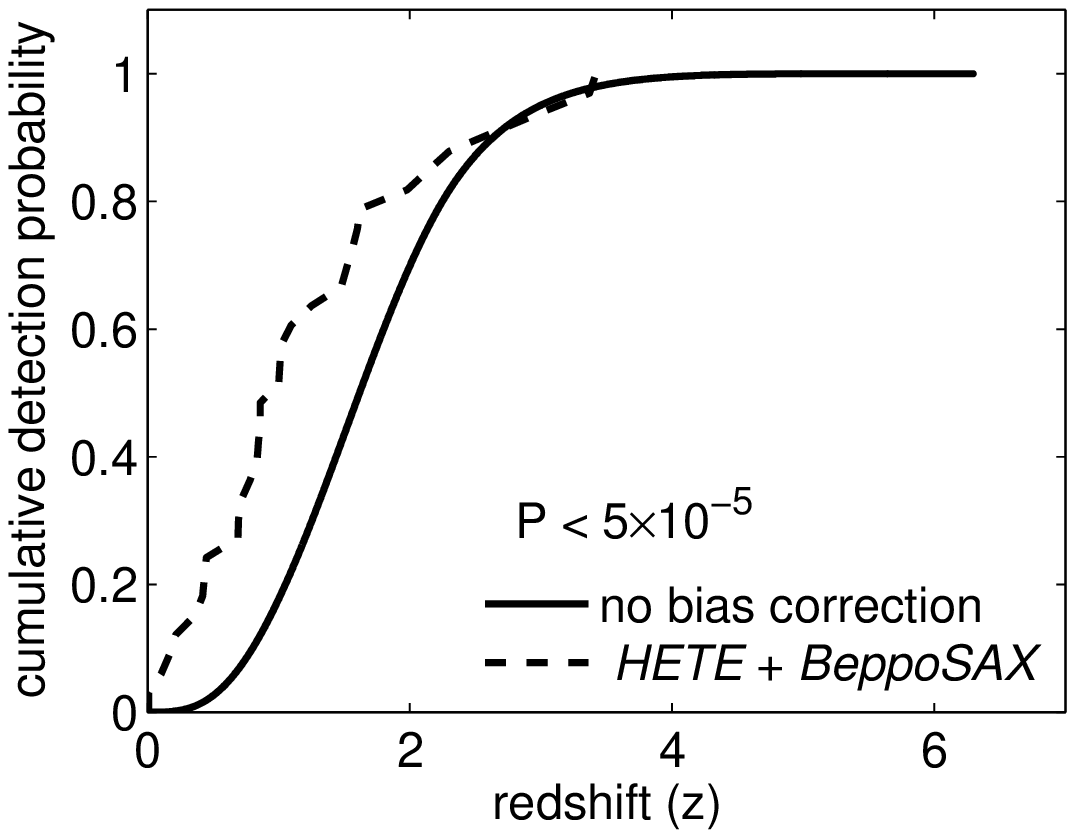}
\includegraphics[scale=0.65]{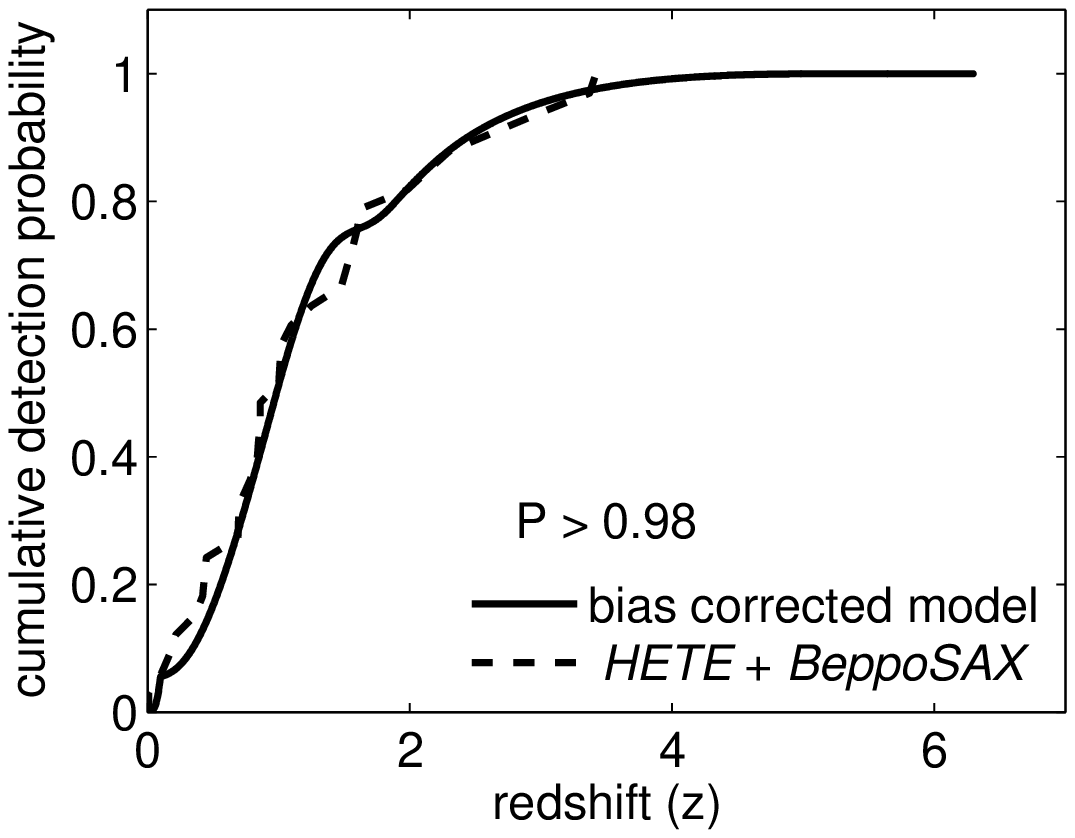}
\caption{Same as figure 3, but using the {\it HETE}+{\it BeppoSAX} redshift
distribution with the same optimized parameter values shown
above. To account for the relatively longer average time delay
in optical afterglow follow-up compared to {\it Swift}, we include a
suppression factor at $z>1$ of $z^{-\delta}$. The optimal KS probability
constrains $\delta$ to $0.4$.} \label{fighb} 
\end{figure}

We apply the weighting technique to the {\it Swift} redshift distribution
using the parametrized model above to constrain the fitting parameters in the redshift ranges $z = 0.1-1.85$ and $z > 1.85$. At the 90\% KS probability level, the free parameters are constrained to ($R_{LL}$, $\alpha$, $\beta$, $\gamma$) = (20-90, 1.3-1.8, 0.35-0.45, 1.45-1.55).  Cumulative distributions of the {\it Swift} redshift distribution and a bias correction model
are plotted in figure \ref{figsw} and \ref{fighb}. We find the highest KS probability of 98\%
using ($R_{LL}$, $\alpha$, $\beta$, $\gamma$) = (40, 1.7, 0.38, 1.5). For the {\it HETE}+{\it BeppoSAX} sample we find similar values for the above parameters, except for $R_{LL} \approx 30$, slightly smaller than that obtained using the {\it Swift} distribution--see figure 4. To be compatible with the {\it HETE}+{\it BeppoSAX} distribution, the bias model has to include the additional suppression of detections at $z>1$ with $\delta=0.4$. 

In the volume bounded by $z \approx 0.1$ for the {\it Swift} and {\it HETE}+{\it BeppoSAX} distributions, the optimal KS probabilities imply a rate increase of $20-90$ times that of the higher-$z$ HL GRBs. This could be evidence for a separate GRB population with a luminosity distribution biased towards much lower luminosities, compared to the HL population--see \cite{Cow05} for an early attempt at modelling this population and \cite{liang07} for more detailed analysis. Alternatively, the LL GRBs may represent a component of a single population characterized by nearly isotropic emissions. 

Here, we assume a single LF encompasses all GRB luminosities. The relative increase of the GRB rate in small $z$ may be attributed to two factors: a larger mean GRB jet opening angle, $\theta$, and the small volume that the GRB detectors are sensitive to. The beaming factor, $f^{-1}_{LL,HL}(\theta)=1-$cos $\theta$, defines the number  of GRBs not observed for every one that is observed. We can estimate $\theta$ for the nearby GRBs via $f^{-1}_{LL}(\theta) = f^{-1}_{HL}(\theta)/ R_{LL} $, where $f^{-1}_{LL}(\theta)$ and $f^{-1}_{HL}(\theta)$ are the beaming factors for the LL and HL GRBs respectively. Using $R_{LL} \sim 40$ and an estimate of $10 \degr$ for the HL GRB beaming angle \citep{guetta05}, we obtain $\theta_{LL} \sim 70 \degr$. This is compatible with observations of the late-time radio afterglow of GRB/XRF 060218, with a jet opening angle that appeared to be much less collimated \citep{sod06}.
 
 Using the bias correction model $\psi_{\rm redshift}(z)$ with ($\alpha$, $\beta$, $\gamma$) = (1.7, 0.38,  1.5) and equation (\ref{fdark}), we calculate the fraction of unobserved redshifts for GRBs localized by {\it Swift}. We find that 72\% of GRBs redshifts are missing at $z < 2$, 55\% at $z > 2$ and a total fraction $\approx 60\%$ over all redshift. This result is independent of the physical origin of the biases.

\begin{figure} 
\includegraphics[scale=0.67]{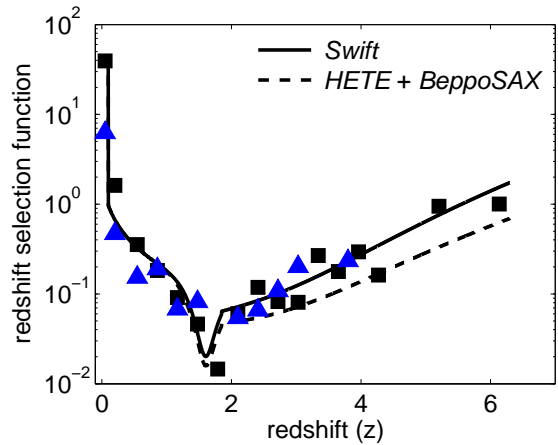}
\caption{Plots of the most probable redshift detection bias functions for GRBs localized by {\it Swift} and {\it HETE}+{\it BeppoSAX} using $\psi_{\rm redshift}$ with $(R_{LL},\alpha,\beta,\gamma)=(40, 1.7, 0.38,1.5)$. The normalized bias function is the same for {\it Swift} and {\it HETE}+{\it BeppoSAX} apart from the power law with $\delta = 0.4$ at $z>1$ for the latter. Data, squares for {\it Swift} and triangles for {\it HETE}+{\it BeppoSAX}, are obtained from weighting the distributions by $\psi_{\rm flux}(z)e(z)(1+z)^{-1}dV/dz$.  The models include an enhanced rate density of $40$ at $z<0.1$ to achieve a maximum KS probability $>98\%$. } \label{fig_select}
\end{figure}

Figure \ref{fig_select} reveals some interesting trends in optical afterglow and redshift non-detection, as a function of redshift.  It is clear that the strongest evolution of optical afterglow extinction is occurring at $z=0-1$, outside of the redshift desert at $z\approx1.5-2$. For $z>2$, the effect of the bias is reducing with redshift, implying that the probability of measuring a redshift is increasing with redshift.

\section{Discussion}
We show that as the total rate of GRBs increases with the SFR at $z  =0 -1$, the observed fraction of GRB redshifts decreases at a similar rate. This is a surprising result, and in effect implies that GRB rate evolution, assuming it follows the SFR,  cancels from the distribution model. The fitting results for $z<1$ imply that if GRB rate evolution is linked to the SFR, there must be a bias that compensates for this rate increase. One possibility is that GRB redshift determination becomes more difficult in redshift ranges where star formation is prolific. It is difficult at present to untangle the relationship between the SFR and dust extinction of GRB afterglows. But we do know that the SFR evolves strongly at $z = 0 - 1$, the range where most of the GRB redshifts are missing.  Using the most probable model for $\psi_{\rm redshift}(z)$, $F_{dark}(z)$ yields a total fraction of missing GRB redshifts $\approx 63\%$, a result supported independently by the observed ratio of GRB redshifts to GRBs of about $70\%$. Given that 55\% of GRB optical afterglows are extinguished by host galaxy dust and absorption at high redshift \citep{rom2006}, our results are suggestive of an extinction function that may compensate for GRB rate evolution tracking the SFR.

An inspection of figure \ref{fig_select} at high $z$ presents a puzzle. If the dominant biases that modify the redshift distribution are distance dependant, such as dust extinction, then why is the probability of observing an afterglow and measuring a redshift increasing with redshift. Several speculative ideas that are compatible with this result include:
\begin{enumerate}
\item GRBs are increasing in luminosity at high redshifts, resulting in a bias for selecting optical afterglows at $z>2$. 
\item the GRB rate at $z >2$ is evolving faster than the assumed global SFR in the same redshift range. 
\item an evolution in the type of dust-i.e. evolving opacity at high redshift. 
\end{enumerate}
The first suggestion has been proposed before to explain the relatively large number of highly luminous GRBs observed at high $z$. We note that if we assume that the GRB rate does not follow the global SFR, but increases more gradually, the strong bias at $z=0.1-1$ is reduced. This is obvious from our analysis, since GRB rate evolution that follows the global SFR is essentially cancelled out by the selection function. But we cannot rule out that the possibility that the global SFR model used in this study underestimates the `true' rate at high $z$. 

Finally, our analysis supports other work finding increased GRB numbers at small $z$, corresponding to LL GRBs. For both the {\it Swift} and {\it HETE}+{\it BeppoSAX} distributions we find a similar increased rate of GRBs is required at small redshift compared with high-$z$ bursts. If the enhanced rate results from a nearly isotropic emission component, the rates of HL and LL GRBs are similar \citep{guetta07}.  

The challenge of understanding the dominant bias from the
GRB redshift sample can be tackled in two ways. Firstly,
small number statistics result in significant uncertainty in
the parameters that comprise $\psi_{\rm redshift}(z)$.
As the number of {\it Swift} redshifts approaches one hundred, the
shape of  $\psi_{\rm redshift}(z)$ at $z = 0-1$ should be further constrained
to the point where a formal fit to the data becomes
more useful. Secondly, to achieve this goal requires
constructing models for  $\psi_{\rm redshift}(z)$ that relate to the physical
processes that cause the bias. A significant improvement to the heuristic approach used here would be to implement a universal extinction function for GRB afterglows. This would provide a test of whether such a model is reconcilable with the evolution of biases reported here. 

It is especially critical to untangle the relationship between
dust extinction of the optical afterglow, GRB redshift determination, GRB rate evolution and the SFR.
Is the reduction in redshift numbers observed at $z  = 0-1$
somehow related to the SFR, or does this suppression originate from an entirely
different class of biases? An ironic possibility that this work
highlights is that the population of GRBs without measured redshifts may tell us just as much about GRB environments as the population with redshifts.

\section*{Acknowledgments}

The authors thank the referee for providing invaluable comments and suggestions that have significantly improved the presentation of this study.  D.M. Coward is supported by the Australian Research Council grant LP0667494. He thanks the organizers of the workshop `The Next Decade of Gamma-ray Burst Afterglows -- Amsterdam 2007', for providing a fertile collaborative environment for ideas that led to this work. 
\label{lastpage}


\begin{thebibliography}{99}

\bibitem[Band(2006)]{Band06}Band D., 2006, ApJ, 644, 378

\bibitem[Blain \& Natarajan(2000)]{Blain00}Blain A.W., Natarajan P., 2000, MNRAS, 312, L35

\bibitem[Bloom(2003)]{Bloom03}Bloom J.S., 2003, AJ, 125, 2866

\bibitem[Coward(2005)]{Cow05}Coward D.M., 2005, MNRAS, 360,  L77-L81

\bibitem[Coward(2007)]{Cow07}Coward D., 2007, New Astron. Rev., 51 539

\bibitem[Fiore et al.(2007)]{fiore07}Fiore F., Guetta G., Piranomonte S., Elia V.D.Õ, Antonelli L.A., 2007, A\&A, 470, 515

\bibitem[Fruchter et al.(2006)]{fruch06}Fruchter A., et al., 2006, Nature, 441, 463

\bibitem[Guetta, Piran \& Waxman(2005)]{guetta05} Guetta D., Piran T.,  Waxman E., 2005, ApJ, 619, 412

\bibitem[Guetta \& Della Valle(2007)]{guetta07} Guetta D., Della Valle M., 2007, ApJ, 657, 73


\bibitem[Gou et al.(2004)]{gou04}Gou L.J., Meszaros P., Abel T., Zhang B., 2004, ApJ, 604, 508

\bibitem[Hjorth et al.(2003)]{hjorth03}Hjorth J., et al., 2003, Nature, 423, 847

\bibitem[Hopkins \& Beacom(2006)]{hb06} Hopkins A.M., Beacom J.F., 2006, ApJ, 651, 142

\bibitem[Jakobsson et al.(2006)]{jakob06}Jakobsson et al., 2006, A\&A, 447, 897

\bibitem[Le FlocÕh et al.(2006)]{FlocÕh06}Le Floc'h E., Charmandaris V., Forrest W.J., Mirabel I.F., Armus L., Devost D., 2006, ApJ, 642, 636

\bibitem[Levan et al.(2006)]{lev06}Levan A., et al., ApJ, 2006, 647, 471

\bibitem[Liang  et al.(2007)]{liang07}Liang E., Zhang B., Virgili F., Dai Z.G., 2007, 662, 1111



\bibitem[Meszaros et al.(2005)]{Mz05}Meszaros P., Bagoly Z., Klose S., Ryde F., Larsson S.,
Balazs L.G., Horvath I., Borgonovo L., 2005, Nuovo
Cim., 28C, 311

\bibitem[Mirabel \& Rodriques(2003)]{MR03}Mirabel I.F., Rodriques I., 2003, Science, 300, 1119





 \bibitem[Roming et al.(2006)]{rom2006}Roming P.W.A. et. al., 2006, ApJ, 652, 1416
 
   \bibitem[Schady et al.(2007)]{schad2007}Schady P. et al., MNRAS, 2007, 377, 273
   
   \bibitem[Soderberg et al.(2006)]{sod06}Soderberg A.M., et al., 2006, Nat. 442, 1014
  
  \bibitem[Stanek et al.(2003)]{stan03}Stanek S. K., et al. 2003, ApJ, 591, L17

\bibitem[Tanvir \& Jakobsson(2007)]{T2007}Tanvir N.R., Jakobsson P., 2007, Phil. Trans. Royal Society A, 366, 1377, arXiv:astro-ph/0701777

 \bibitem[Woosley(1993)]{woos93}Woosley S. E., 1993, ApJ, 405, 273
 
 \bibitem[Yoon \& Langer (2005)]{yoon05}Yoon S. C., Langer N., 2005, A\&A, 443, 643





\end{thebibliography}
\end{document}